# Negative resistance state in superconducting $NbSe_2$ induced by surface acoustic waves

**Masahiko Yokoi[1], Satoshi Fujiwara[1], Tomoya Kawamura[1], Tomonori Arakawa[1,2],**

**Kazushi Aoyama[3], Hiroshi Fukuyama[4,5], Kensuke Kobayashi[1,4,6], and Yasuhiro Niimi[1,2,*]**

[1]Department of Physics, Graduate School of Science, Osaka University, Toyonaka 560-0043, Japan.

[2]Center for Spin Research Network, Osaka University, Toyonaka 560-8531, Japan.

[3]Department of Earth and Space Science, Graduate School of Science, Osaka University, Toyonaka 560-0043, Japan.

[4]Department of Physics, Graduate School of Science, The University of Tokyo, Tokyo 113-0033, Japan.

[5]Cryogenic Research Center, The University of Tokyo, Tokyo 113-0032, Japan.

[6]Institute for Physics of Intelligence, Graduate School of Science, The University of Tokyo, Tokyo 113-0033, Japan.

*To whom correspondence should be addressed. E-mail: niimi@phys.sci.osaka-u.ac.jp (Y. N.)

**We report a negative resistance, namely, a voltage drop along the opposite direction of a current flow, in the superconducting gap of $NbSe_2$ thin films under the irradiation of surface acoustic waves (SAWs). The amplitude of the negative resistance becomes larger by increasing the SAW power and decreasing temperature. As one possible scenario, we propose that soliton-antisoliton pairs in the charge density wave of $NbSe_2$ modulated by the SAW serve as a time-dependent capacitance in the superconducting state, leading to the *dc* negative resistance. The**

**present experimental result would provide a previously unexplored way to examine nonequilibrium manipulation of the superconductivity.**

**Introduction**

Negative differential resistances in semiconductors are well-established in tunnel diodes (or Esaki diodes) (*1*) and Gunn diodes (*2*), which are used in electronic oscillators and amplifiers, particularly at microwave frequencies. In those cases, the negative differential resistances are realized at finite bias regions. In other words, the absolute value of resistance, that is a voltage $V$ divided by an applied current $I$, remains positive. In general, a negative resistance state at the zero bias limit violates either the conservation law of energy or the second law of thermodynamics, and thus is unstable in normal electric circuits (*3*). By injecting an additional power such as microwaves from the outside, however, electronic transport properties can dramatically be changed. A typical example is microwave-induced zero resistance in a two-dimensional (2D) electron system (2DES) (*3-7*). This nonequilibrium phenomenon occurs when the microwave frequency coincides with the Landau level splitting energy in the 2DES in a perpendicular magnetic field (*4-7*). While a negative resistance can be realized in a finite-size normal electron system at nonequilibrium (*8*), it is not a steady state in the thermodynamic limit as mentioned above, resulting in the zero resistance state (*3,7*).

In a superconducting state where the resistance is zero up to the critical current $I_C$, Ohm's law is not valid but instead the London equation is used to explain the electrodynamic properties (*9*). When microwaves are irradiated to superconducting junctions, voltage jumps can be observed in the *I-V* curve (*9*), that is known as the *ac* Josephson effect. In this case, the resistance is still zero (or slightly positive). Here we consider the situation that superconducting thin films (*10-15*) on a piezoelectric substrate are irradiated by surface acoustic waves (SAWs). The SAW (more precisely Rayleigh wave) is an elastic wave traveling along the surface of piezoelectric substrate and is generated by applying an *ac* voltage to comb-shaped electrodes, as illustrated in Fig. 1A. The excited SAW can be coupled with the lattice system of thin films on the substrate, resulting in the modulation of electronic properties of the thin films (*16-22*).

In this report, we demonstrate that the irradiation of SAWs induces a negative resistance at the zero bias current limit, i.e., an opposite voltage drop with respect to the current direction, in a superconducting gap of thin 2*H*-NbSe$_2$ films. The amplitude of the negative resistance is enhanced with increasing the SAW power and decreasing temperature. The negative resistance does not appear in either 2*H*-NbS$_2$ or Nb devices which have no charge density wave (CDW) phase, but is realized only in the NbSe$_2$ devices where the CDW and the superconductivity coexist below its superconducting critical temperature ($T_C$). This fact strongly indicates that the negative resistance is caused by cooperative interplay between the CDW modulated by the SAW and the superconductivity. To give an insight into the observed negative resistance, we propose a possible model where the negative resistance can be explained by taking into account time-dependent soliton-antisoliton pairs induced by the SAW and the Josephson junction in series.

**Results**

2*H*-NbSe$_2$ is a typical transition metal dicalcogenide showing the superconductivity below $T_C$ = 7.2 K. By using the mechanical exfoliation technique, NbSe$_2$ thin films were fabricated on a LiNbO$_3$ substrate which is a strong piezoelectric material. The NbSe$_2$ film selected in the present work is about 30 nm in thickness, which corresponds to 25 unit cells. In order to induce SAWs along the *x*-direction shown in Fig. 1A and to apply them to the NbSe$_2$ thin film on the LiNbO$_3$ substrate, we placed two opposite interdigital transducers (IDTs) so that the selected thin film was at the center of the two IDTs, as shown in Fig. 1B. The electrodes and the IDTs were prepared at the same time using electron beam lithography and depositions of Ti (60 nm) and Au (40 nm) in a vacuum chamber. Each IDT consists of 10 pairs of comb-shaped electrodes. The wavelength and the resonance frequency ($f_0$) for the designed IDTs are 1.0 μm and 3.25 GHz, respectively. The latter has been determined from the scattering (*S*)-parameter measurement with a network analyzer. The amplitude of the induced SAW at $f_0$ can be estimated at about 0.3 Å when the radio frequency (*rf*) power (or SAW power) applied to the IDTs is 30 μW (*23*). As reference samples, we also prepared 30 nm thick 2*H*-NbS$_2$ and Nb thin films on other LiNbO$_3$ substrates.

Figure 1C shows the temperature dependence of resistance, measured with the standard *ac* lock-in technique, for a typical NbSe$_2$ thin film on the LiNbO$_3$ substrate. A broad kink due to the CDW transition and a sharp superconducting transition can be seen at 33 K and 7.2 K respectively, as well-established for bulk NbSe$_2$ (*24,25*). A similar NbS$_2$ thin film device also shows a superconducting transition at 4 K (see Fig. 1C). In contrast, it has no CDW phase (*25*), which is supported by the observed featureless temperature dependence of resistance.

In Figs. 2A and 2B, we show *dc* current-voltage (*I-V*) properties of the NbSe$_2$ film exposed to the SAW sent from the left to right IDTs (see Fig. 1B) with different *rf* powers. The standard *I-V* curve is obtained except for in the vicinity of zero current. The critical current $I_C$ for this device is about 0.8 mA, which is almost independent of the *rf* power (up to 30 μW). This shows that the superconducting state of the $NbSe_2$ device is robust for the irradiation of SAW. The shape of the differential resistance *dV/dI* just above $I_C$ slightly changes with increasing the power, as depicted in Fig. 2C. This would be caused by the heating effect of the device.

An unexpected property emerges in the *I-V* curve within ±50 μA (see Fig. 2B). When the *rf* power is as small as 1 μW, zero voltage is detected within the critical current. Surprisingly, however, a finite voltage appears at ±15 μA and the voltage amplitude increases with increasing the *rf* power. Since the negative (positive) voltage is induced at +(−)15 μA, the slope, namely resistance at $I = 0$ is negative. This fact indicates that the induced *dc* voltage drop is opposite to the current direction. To see the negative resistance more clearly, the differential resistance *dV/dI* numerically obtained from the *dc I-V* curve is plotted in Fig. 2D as a function of *I* for different *rf* powers. The negative differential resistance can be seen in the *I* regions of ±15 μA and also from +(−)20 to +(−)30 μA. At $T = 1.6$ K and the *rf* power of 30 μW, the resistance of $NbSe_2$ is as large as −0.16 Ω, comparable to that (+0.3 Ω) of the normal state. A similar negative differential resistance was also observed when the SAW direction was inverted, as shown in Supplementary Materials (see Fig. S1C). On the other hand, such a negative differential resistance has never been observed in the $NbS_2$ (see Figs. 2E-2H) and Nb devices (see Fig. S4 in Supplementary Materials). This fact suggests the relevance of the CDW phase for the negative resistance.

The negative resistance is further investigated by measuring the temperature dependence and the magnetic field dependence. Figures 3A and 3B show the resistance of the NbSe$_2$ device obtained by the *ac* lock-in technique as a function of temperature and magnetic field, respectively. We note that the resistance obtained with the *ac* method corresponds to the differential resistance of the *I-V* curve in the zero bias limit and enables us to continuously sweep as a function of temperature. Although the critical temperature $T_C$ (= 7.2 K) and critical field $H_C$ (= 3.3 T) are independent of the *rf* power, the negative resistance of the NbSe$_2$ device becomes smaller with increasing temperature and magnetic field. In contrast, the resistance of the NbS$_2$ device is zero up to $T_C$ (= 4 K) and $H_C$ (= 0.8 T). We have confirmed the above tendencies for four different NbSe$_2$ and two different NbS$_2$ devices from different batches.

**Discussion**

What is the origin of the negative resistance in the superconducting gap observed only in the NbSe$_2$ devices? The negative resistance takes place without the external magnetic field. Thus, we can exclude microwave-induced resistance oscillations (*3-8*) mentioned in the introduction. Recently, a negative local resistance has been reported in graphene where the high viscosity of electronic states is essential (*26*), but such high viscosity cannot be expected in the NbSe$_2$ devices since the electron density in NbSe$_2$ is much higher than that in graphene. Although it is known that a very large contact resistance may induce a seeming negative resistance (*27*), this possibility can be ruled out safely from the following two reasons: firstly, the negative resistance is observed only when the SAW is applied to the NbSe$_2$ device, and is never observed for the NbS$_2$ device that has the same contact resistance (~100 Ω) as the NbSe$_2$ one; secondly the *I-V* characteristic above $T_C$ is perfectly Ohmic. We can also exclude the possibility that the sufficiently inhomogeneous current distribution (most probably due to the CDW phase as detailed in the next paragraph) in the NbSe$_2$ devices might induce a negative resistance. This is because the negative resistance does not appear just below $T_C$ where the coexistence of the CDW and the superconductivity is already realized, but appears only far below $T_C$ (see Fig. 3A). This fact provides assurance that the current density is relatively homogeneous in the NbSe$_2$ devices. The emergence of a similar negative resistance was already reported in Josephson junctions (*28-30*). In Refs. (*28*) and (*29*), excess quasiparticles excited outside the superconducting gap tunnel opposite to the current bias direction due to the asymmetric profile of the density of states, leading to a negative resistance. In that case, however, the zero resistance state is not recovered again at finite bias regions, which is different from the present case (see Figs. 2 and S5D). In Ref. (*30*), a negative resistance state was obtained only when the amplitude of *rf* current was much larger than the *dc* critical current of the Josephson junction. This situation is also different from the present setup where the SAW power is small enough not to affect $I_C$, $T_C$, and $H_C$ of NbSe$_2$ thin films, as demonstrated in Figs. 2 and 3.

Therefore, it is reasonable to attribute the present negative resistance to cooperative interactions between the superconducting state and the CDW modulated by the SAW. At the moment, we do not have a conclusive model to fully explain our experimental results, but one possible scenario is that soliton-antisoliton pairs in the CDW phase (*31-33*) are generated by irradiating the SAW and they form local capacitances in the superconducting domains. In the CDW phase of NbSe$_2$ (below 33 K), selenium atoms have a periodic modulation that is three times the lattice constant for selenium atoms (*24,34,35*). When the SAW is irradiated to the NbSe$_2$ device, all the selenium atoms would be shifted from the commensurate position of the CDW and thus the phase $\varphi$ of the CDW order parameter would be modulated over the wavelength of the SAW (= 1 μm), as shown in the upper panel of Fig. 4A. This displacement increases the electrostatic energy as well as the elastic energy. Thus, it is energetically more favorable to nucleate soliton-antisoliton pairs and to induce a phase difference $\Delta\varphi$ of $2\pi$ in between the soliton-antisoliton pairs (*31-33,36*), as illustrated in the lower panel of Fig. 4A. Such soliton-antisoliton pairs have been intensively studied in quasi-one-dimensional CDW systems (*31,32*) and also discussed even in 2D CDW compounds (*33*). Very recently, it has been revealed that there are several types of domains with a size of about 10 nm in the equilibrium NbSe$_2$ (*35*). This is an additional supportive observation for the soliton-antisoliton pairs in the NbSe$_2$ film. In addition to the complex CDW order, the superconductivity starts to develop below 7.2 K. Here we assume the following two situations: 1) a superconducting domain described by a macroscopic wave function grows on each CDW domain with a size of ~10 nm (see Ref. 35), and 2) boundaries of the superconducting domains randomly distributed in the NbSe$_2$ film serve as weak junctions.

While the CDW opens an energy gap at the Fermi energy, it is not a full-gap state for NbSe$_2$ (*37*) and thus the charge accumulation due to the soliton-antisoliton pair creation is immediately dissolved. On the other hand, the superconducting gap fully opens at the Fermi energy below $T_C$. Therefore, the charge accumulation is expected to survive longer in the superconducting state so that it can be regarded as a temporal and local capacitance. As for the superconducting part, the shape of the *I-V* curve near $I_C$ is typical of an overdamped Josephson junction because there is no hysteresis near $I_C$ (*9*). This fact indicates the existence of weakly coupled superconducting domains in the NbSe$_2$ thin film, which is also consistent with the above assumption that the superconducting parts developed on different CDW domains (35) are weakly coupled.

Based on the above experimental facts and ideas, we have developed a capacitively coupled Josephson junction model (see Supplementary Materials for more details), and calculated an *I-V* property of the circuit shown in the upper panel of Fig. 4B. In this calculation, we assume that the time(*t*)-dependent local capacitance $C(t)$ between the soliton-antisoliton pairs is modulated by the same frequency as the SAW to have a sawtooth wave as shown in the lower panel of Fig. 4B. A typical *I-V* curve based on this model is displayed in Fig. 4C (see Supplementary Materials for derivation). We qualitatively reproduce a negative resistance in the superconducting gap; the negative slope appears at zero current in the superconducting gap.

On the other hand, there is a difference between the experimental result and the theoretical calculation; in most cases (see Figs. 2B and S5D in Supplementary Materials), the negative resistance is realized only in the vicinity of the zero current and the zero resistance state is recovered again at finite bias regions (up to $I_C$), while in the present calculation the negative resistance continues up to $I_C$. In real devices, Josephson junctions should be randomly distributed over the NbSe$_2$ films presumably constituting a network of superconducting domains, and there should be soliton and antisoliton pairs in the films when the SAW is irradiated. However, this is simplified by the one-dimensional model in the present theoretical model. In addition, the dynamics of soliton-antisolition pairs is also simply modeled as the time-dependent capacitance in this work, but it could be affected by the supercurrent flow via the microscopic interplay between the CDW and superconductivity (*38*). The effect of such an interplay should become relevant for larger supercurrents. The above issues are the future works to be resolved.

In conclusion, we have demonstrated a negative resistance in the superconducting gap of NbSe$_2$ thin film on the LiNbO$_3$ substrate induced by the SAW irradiation. Our experiment indicates that it occurs due to the interplay between the superconductivity and the SAW-modulated CDW. By using a theoretical model based on the capacitively coupled Josephson junction under the SAW irradiation, we have qualitatively reproduced the experimentally measured negative resistance. Such a negative resistance state could be a promising stage to demonstrate Floquet engineering (*39*) in the superconducting state where quantum systems can be driven by the SAW.

**Materials and Methods:**

2$H$-NbSe$_2$ and 2$H$-NbS$_2$ thin films have been obtained using the mechanical exfoliation technique using scotch tape. Since these thin films are sensitive to ambient air, the exfoliation process has been performed in a globe box filled with Ar gas of purity 99.9999%. Some of the exfoliated NbSe$_2$ (or NbS$_2$) flakes were transferred from the scotch tape to a LiNbO$_3$ substrate. We then spin-coated polymethyl-methacrylate (PMMA) resist onto the substrate for electron beam lithography and also for protecting the thin films from degradation. The substrate was taken out from the glove box, and electrode patterns including comb-shaped electrodes for irradiation of SAWs were printed using electron beam lithography. After the lithography, the substrate was put back into the glove box again for the development of the resist. The Ti (60 nm) and Au (40 nm) electrodes were deposited using electron beam deposition in a vacuum chamber next to the glove box. The NbSe$_2$ and NbS$_2$ thin film SAW devices can be obtained after the lift-off process in acetone. For a reference sample, we also prepared Nb thin film SAW-devices using magnetron sputtering.

Low temperature transport measurements of the SAW devices were performed using a cryogen-free superconducting magnet system (Cryomagnetics, C-mag Vari-9) down to $T$ = 1.6 K and up to $B$ = 9 T. In *dc* measurements, the bias current was generated by using a voltage source (Yokogawa, 7651) and a large resistance (~ 1 MΩ). The obtained *dc* voltage was measured by a digital Multimeter (Keythley, 2000). We also performed *ac* measurements with a frequency of 173 Hz using a Lock-in amplifier (Stanford Research System, SR830) to obtain a differential resistance. The fundamental properties of the IDTs were confirmed by measuring *S*-parameters using a network analyzer (Keysight, E5071C). The SAWs were induced by a microwave generated from a signal generator (Agilent, N5171B).

**Supplementary Materials:**

Supplementary Text

Figs. S1 to S5

**Acknowledgements**

We would like to thank C. Bäuerle, H. Shimada, N. Takeuchi, S. Kawabata, T. Kato, S. Takayoshi, and T. Oka for fruitful discussions. We acknowledge the stimulated discussion in the meeting of the Cooperative Research Project of RIEC, Tohoku University. This work was supported by JSPS KAKENHI (Grant Numbers JP16H05964, JP17K18756, JP19K21850, JP26103002, and JP26220711), the Mazda Foundation, Shimadzu Science Foundation, Yazaki Memorial Foundation for Science and Technology, SCAT Foundation, and the Murata Science Foundation, Toyota Riken Scholar, and Kato Foundation for Promotion of Science.

**Author Contributions:**

M. Y., T. A., and Y. N. designed the experiments. M. Y., S. F., and T. K. fabricated devices and performed the experiments. H. F. provided bulk NbSe$_2$ samples. K. A. performed the theoretical calculations. M. Y., K. A., K. K., and Y. N. wrote the manuscript.

All the authors discussed the results and commented on the manuscript.

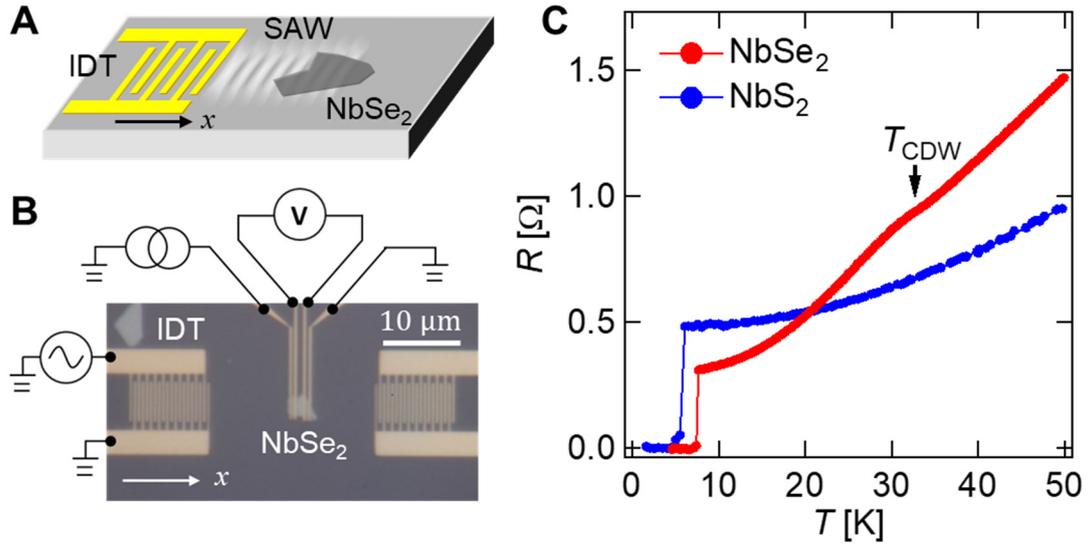

**Fig. 1**. **Experimental setup and fundamental properties of the device. (A)** Schematic image of the device structure. The SAW (white wave) emitted from comb-shaped electrodes travels along the *x*-direction and irradiates a NbSe$_2$ thin film (dark gray flake). **(B)** Optical microscope image of the device and schematic image of the circuit. The NbSe$_2$ film, whose thickness is about 30 nm, is attached on a LiNbO$_3$ substrate. The film is electrically connected by Ti/Au electrodes. Each IDT consists of 10 pairs of electrodes. The resonance frequency of the IDTs is 3.25 GHz. **(C)** Temperature dependence of resistances of the NbSe$_2$ and NbS$_2$ devices measured with the four-terminal lock-in technique. The bias current is 1 μA. A small bump due to the CDW phase is observed at 33 K only for the NbSe$_2$ device.

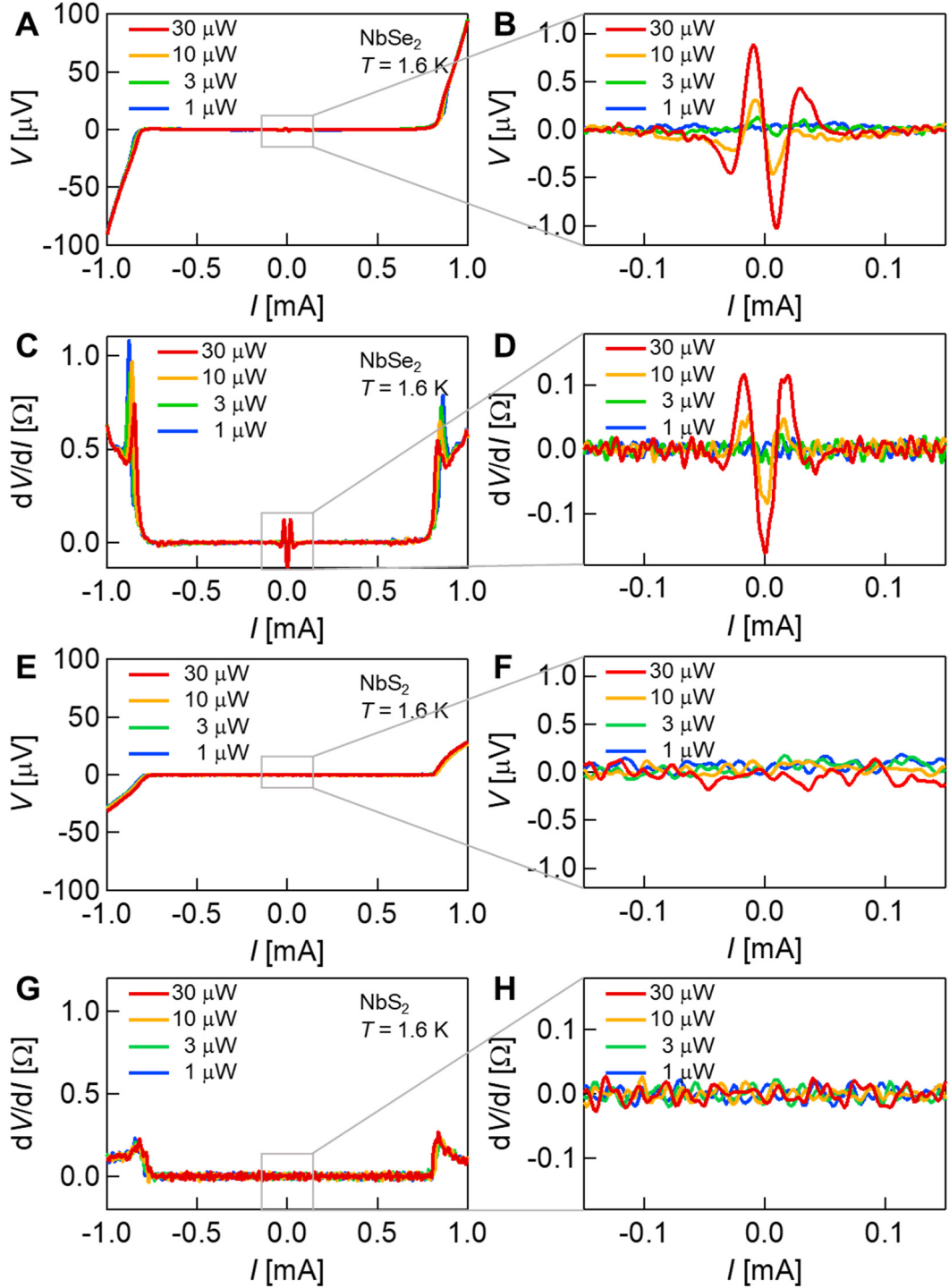

**Fig. 2. Current-voltage characteristics of NbSe$_2$ and NbS$_2$ devices exposed to SAWs.** (**A**) *Dc* voltage of the NbSe$_2$ thin film exposed to the SAW with different *rf* powers as a function of bias current measured at 1.6 K. (**B**) Closeup of the *I-V* curve

shown in (A) near zero bias current. **(C)** Differential resistance obtained by numerically differentiating the *I-V* curve shown in (A). **(D)** Closeup of the *dV/dI* curve shown in (C) near zero bias current. **(E)** *Dc* voltage of the NbS$_2$ thin film exposed to the SAW with different *rf* powers as a function of bias current measured at 1.6 K. **(F)** Closeup of the *I-V* curve shown in (E) near zero bias current. **(G)** Differential resistance obtained by numerically differentiating the *I-V* curve shown in (E). **(H)** Closeup of the *dV/dI* curve shown in (G) near zero bias current.

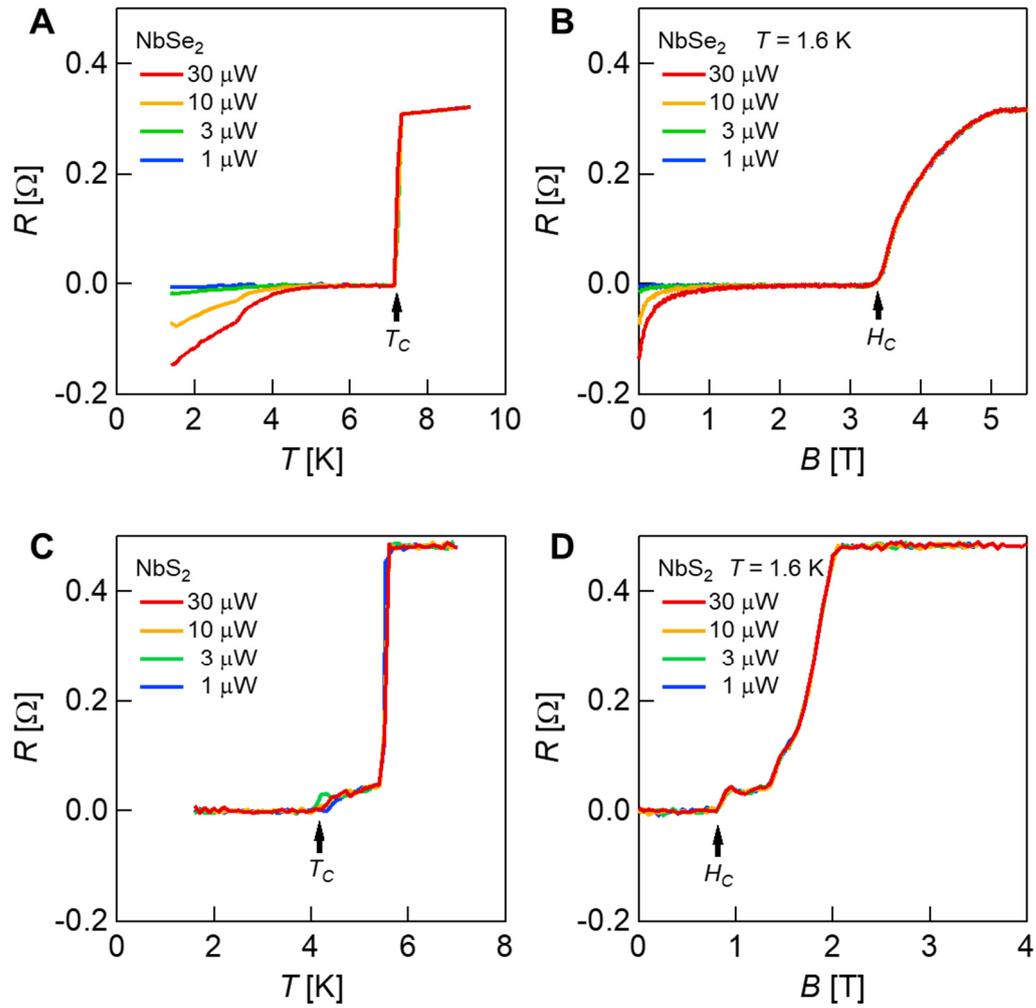

**Fig. 3. Temperature and magnetic field dependencies of resistances of NbSe$_2$ and NbS$_2$ devices exposed to SAWs. (A)** Temperature dependence of resistance of the NbSe$_2$ device with different SAW powers. **(B)** Magnetic field dependence of resistance of the NbSe$_2$ device measured at 1.6 K and with different SAW powers. **(C)** Temperature dependence of resistance of the NbS$_2$ device with different SAW powers. **(D)** Magnetic field dependence of resistance of the NbS$_2$ device measured at 1.6 K and with different SAW powers.

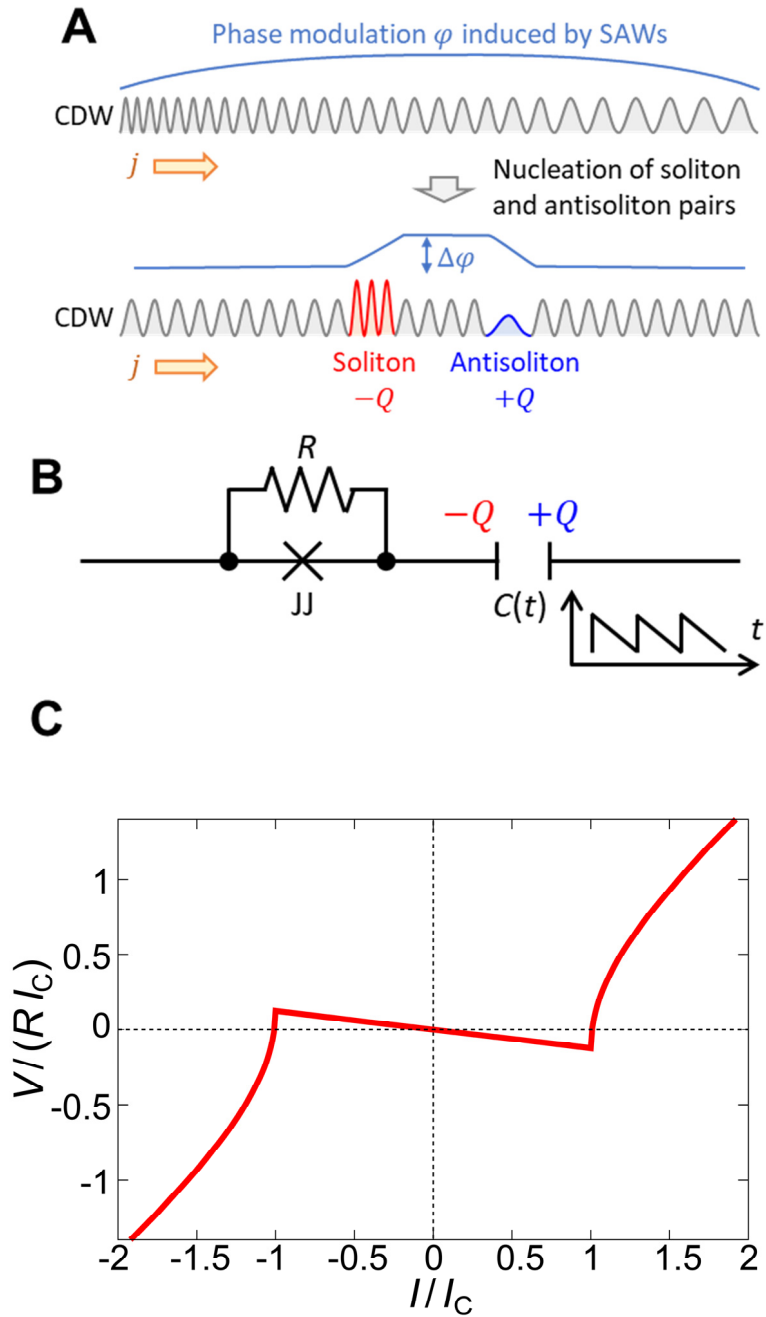

**Fig.4. Theoretical model to reproduce the negative resistance. (A)** Schematics of nucleation of soliton and antisoliton pairs. When the SAW is irradiated to the CDW state, the CDW phase $\varphi$ is modulated over the wavelength of the SAW. Instead of mod-

ulating the CDW over the wavelength, it is more stable to nucleate soliton and antisoliton pairs in the CDW state. In the superconducting state, local charges ($-Q$, $+Q$) are accumulated, resulting in a temporal and local capacitance $C(t)$ and a phase difference $\Delta\varphi$ of $2\pi$. **(B)** The circuit model used in the calculation (see Supplementary Materials for more details). We assume a resistively shunted Josephson junction (JJ) which is capaticively coupled via $C(t)$. The change in $C(t)$ is synchronized with the frequency of the SAW. We assume a sawtooth wave function for $C(t)$. **(C)** A typical *I-V* curve calculated with the above capacitively coupled Josephson junction model. The horizontal and vertical axes are normalized by $I_C$ and the product of the resistance $R$ of the normal state and $I_C$, respectively.